\documentclass[letterpaper,10pt]{article} 

\usepackage{opticameet3} %

\usepackage{amsmath,amssymb}
\usepackage[colorlinks=true,bookmarks=false,citecolor=blue,urlcolor=blue]{hyperref} %

\usepackage{mathtools,amssymb,tikz,bm,siunitx, booktabs, subcaption}
\usetikzlibrary{positioning, shapes.geometric}            %
\usepackage{pgfplots}
\usetikzlibrary{arrows,calc,fit,matrix,positioning,shapes,shadows,trees,mindmap,tikzmark,arrows.meta,angles,quotes,babel,spy}
\usetikzlibrary{external}

\definecolor{kit-green100}{rgb}{0,.59,.51}
\definecolor{kit-green70}{rgb}{.3,.71,.65}
\definecolor{kit-green50}{rgb}{.50,.79,.75}
\definecolor{kit-green30}{rgb}{.69,.87,.85}
\definecolor{kit-green15}{rgb}{.85,.93,.93}
\definecolor{KITgreen}{rgb}{0,.59,.51}

\definecolor{KITpalegreen}{RGB}{130,190,60}
\colorlet{kit-maigreen100}{KITpalegreen}
\colorlet{kit-maigreen70}{KITpalegreen!70}
\colorlet{kit-maigreen50}{KITpalegreen!50}
\colorlet{kit-maigreen30}{KITpalegreen!30}
\colorlet{kit-maigreen15}{KITpalegreen!15}

\definecolor{KITblue}{rgb}{.27,.39,.66}
\definecolor{kit-blue100}{rgb}{.27,.39,.67}
\definecolor{kit-blue70}{rgb}{.49,.57,.76}
\definecolor{kit-blue50}{rgb}{.64,.69,.83}
\definecolor{kit-blue30}{rgb}{.78,.82,.9}
\definecolor{kit-blue15}{rgb}{.89,.91,.95}

\definecolor{KITyellow}{rgb}{.98,.89,0}
\definecolor{kit-yellow100}{cmyk}{0,.05,1,0}
\definecolor{kit-yellow70}{cmyk}{0,.035,.7,0}
\definecolor{kit-yellow50}{cmyk}{0,.025,.5,0}
\definecolor{kit-yellow30}{cmyk}{0,.015,.3,0}
\definecolor{kit-yellow15}{cmyk}{0,.0075,.15,0}

\definecolor{KITorange}{rgb}{.87,.60,.10}
\definecolor{kit-orange100}{cmyk}{0,.45,1,0}
\definecolor{kit-orange70}{cmyk}{0,.315,.7,0}
\definecolor{kit-orange50}{cmyk}{0,.225,.5,0}
\definecolor{kit-orange30}{cmyk}{0,.135,.3,0}
\definecolor{kit-orange15}{cmyk}{0,.0675,.15,0}

\definecolor{KITred}{rgb}{.63,.13,.13}
\definecolor{kit-red100}{cmyk}{.25,1,1,0}
\definecolor{kit-red70}{cmyk}{.175,.7,.7,0}
\definecolor{kit-red50}{cmyk}{.125,.5,.5,0}
\definecolor{kit-red30}{cmyk}{.075,.3,.3,0}
\definecolor{kit-red15}{cmyk}{.0375,.15,.15,0}

\definecolor{KITpurple}{RGB}{160,0,120}
\colorlet{kit-purple100}{KITpurple}
\colorlet{kit-purple70}{KITpurple!70}
\colorlet{kit-purple50}{KITpurple!50}
\colorlet{kit-purple30}{KITpurple!30}
\colorlet{kit-purple15}{KITpurple!15}

\definecolor{KITcyanblue}{RGB}{80,170,230}
\colorlet{kit-cyanblue100}{KITcyanblue}
\colorlet{kit-cyanblue70}{KITcyanblue!70}
\colorlet{kit-cyanblue50}{KITcyanblue!50}
\colorlet{kit-cyanblue30}{KITcyanblue!30}
\colorlet{kit-cyanblue15}{KITcyanblue!15}

\usepackage{wrapfig}
\usepackage[font={footnotesize}]{caption}           %
\DeclareMathAlphabet{\mathcal}{OMS}{cmsy}{m}{n}     %

\newcommand\blfootnote[1]{%
  \begingroup
  \renewcommand\thefootnote{}\footnote{#1}%
  \addtocounter{footnote}{-1}%
  \endgroup
}

\begin{document}

\title{Energy-efficient Spiking Neural Network Equalization for IM/DD Systems with Optimized Neural Encoding}

\author{Alexander von Bank, Eike-Manuel Edelmann, and Laurent Schmalen}

\address{\mbox{Communications Engineering Lab, Karlsruhe Institute of Technology, 76187 Karlsruhe, Germany}}
\email{\texttt{alexander.bank@kit.edu, edelmann@kit.edu}} %

\vspace{-3mm}
\begin{abstract}
We propose an energy-efficient equalizer for IM/DD systems based on spiking neural networks. 
We optimize a neural spike encoding that boosts the equalizer's performance while decreasing energy consumption.
\vspace{-0mm}
\end{abstract}

\blfootnote{This work has received funding from the European Research Council (ERC) under the European Union’s Horizon 2020 research and innovation programme (grant agreement No. 101001899).
Parts of this work were carried out in the framework of the CELTIC-NEXT project AI-NET-ANTILLAS (C2019/3-3) funded by the German Federal Ministry of Education and Research (BMBF) (grant agreement 16KIS1316).}

\vspace{-1mm}

\vspace*{-1mm}
\section{Introduction}
\vspace*{-2mm}
Spiking neural networks (SNNs) enable the implementation of powerful machine-learning algorithms using energy-efficient neuromorphic hardware \cite{arnold23jlt}.
SNNs process information by exchanging short pulses (spikes) between their neurons.
This leads to a sparse representation of information and low energy consumption since spikes are only exchanged when information is processed \cite{Auge21}. 
Recent work has shown that SNN-based equalizers are promising candidates for powerful equalizers with low-energy consumption for optical transmissions \cite{boecherer23, vonBank, arnold_soft, arnold23jlt, bansbach}.
For an intensity modulation / direct detection (IM/DD) link suffering from non-linear impairments and chromatic dispersion (CD), an SNN-based equalizer and demapper, which outperforms linear as well as artificial neural network (ANN)-based equalizers was proposed in \cite{arnold_soft}. 
Simulation results of the equalizer proposed in \cite{arnold_soft} have been reproduced using the PyTorch-based SNN deep learning library \texttt{Norse} \cite{norse} and neuromorphic hardware \cite{arnold23jlt}. 
In \cite{boecherer23}, the equalizer of \cite{arnold_soft} is applied to experimental IM/DD data. 
In \cite{bansbach}, we proposed an SNN-based equalizer with decision feedback, which outperforms the approach of \cite{arnold_soft} for an IM/DD link \cite{vonBank}.

The transformation of continuous data into spiking signals is called neural encoding.
For instance, for the IM/DD link, the channel output is encoded and forwarded to the SNN.
In \cite{Auge21}, several neural encoding schemes and their applications are discussed. 
Three basic encoding schemes are \cite{Auge21}: Rate encoding transforms the information in the spike frequency, i.e., number of spikes per time interval; 
In temporal encoding, the timing of the spikes contains the information; 
For population encoding, the information is encoded in the interaction of different neurons.
Depending on the task, encoding schemes differ in noise robustness, accuracy, energy consumption, and hardware requirements.
Therefore,  a crucial task when implementing SNNs is to find an efficient neural encoding.

While the work mentioned above \cite{arnold23jlt,boecherer23,arnold_soft,bansbach,vonBank} focuses on the design and learning of SNNs, the encoding is designed based on empirical knowledge and not further optimized.   
In \cite{arnold_soft}, a log-scale encoding is proposed, which encodes the information in the relative timing of multiple spikes. In \cite{bansbach,vonBank}, ternary encoding is introduced, which encodes information by activating a predefined subgroup of input neurons.

This work proposes a generic neural encoding based on a learnable matrix, which determines the input pattern fed to the SNN's input layer. 
The learnable matrix is furthermore regularized using the $\ell_p$-over-$\ell_q$ regularization of \cite{Cherni20}.
Using the SNN-based equalizer and simulated IM/DD link of \cite{arnold_soft}, we compare the proposed encoding with log-scale encoding of \cite{arnold_soft} and ternary encoding of \cite{bansbach,vonBank}.
We show that the learned encoding with regularization reduces the number of SNN-generated spikes by up to $50\%$, resulting in reduced power consumption while enhancing the system's performance by $\SI{0.3}{\decibel}$.  
The code is available at \url{https://github.com/kit-cel/OptiSpike}.

\newcommand{\e}{\mathrm{e}}
\vspace*{-1mm}
\section{Spiking Neural Network-based Equalizer}
\vspace*{-2mm}
\begin{wrapfigure}{r}{0.45\textwidth}
    \vspace{-0.9cm}
    \begin{center}
        \resizebox{0.45\textwidth}{!}{
            \begin{tikzpicture}[>=latex,thick]
    \def\zlen{0.65cm}
    \def\zdist{0.15cm}

    \node[draw,rectangle,rounded corners,minimum width=3.5cm, minimum height=1cm] (snn) {\large SNN};

    \node[above left=-0.37cm and 0.05cm of snn, draw, rectangle, rounded corners,minimum height=1.55cm] (enc) {\rotatebox{90}{Encoding}};
    \node[left=1.0cm of enc] (y0) {};
    
    \node[draw, right=0.7cm of enc,rounded corners,minimum width=\zlen,minimum height=\zlen] (zy1) {$z^{-1}$};
    \node[draw, right=0.8cm of zy1,rounded corners,minimum width=\zlen,minimum height=\zlen] (zy2) {$z^{-1}$};
    \node[draw, right=\zdist+0.13cm of enc,circle, inner sep=1pt, fill=black] (nf1) {};
    \node[draw, right=\zdist of zy1,circle, inner sep=1pt, fill=black] (nf2) {};
    \node[below right=-0.9cm and 0.1cm of zy2] (yk) {$y\left[k'-\left\lfloor \frac{d_\text{tap}}{2}\right\rfloor\right]$};

    \node[right=0.3cm of snn, draw, rectangle, rounded corners] (argmax) {\rotatebox{90}{arg$\,$max}};
    \node[right=1.80cm of snn, draw, rectangle, rounded corners] (demap) {\rotatebox{90}{Bit Mapper}};
    \node[right=0.9cm of demap] (xk) {};

    \def\yoff{0.43cm}
    \draw[->] (enc) -- (zy1);
    \draw[->] (enc) -- (nf1) -- +(0cm,-\yoff);
    \draw[->] (zy1) -- (nf2) -- +(0cm,-\yoff);
    \draw[dotted] (nf2) -- +(0.3cm,0cm);
    \draw[->] (nf2)+(0.4cm,0cm) -- (zy2);
    \draw[->] (zy2) -| +(0.52cm,-\yoff);
    \draw[->] (y0) -- (enc) node [pos=0.0,above] {$y[k]$};

    \draw[->] (snn) -- (argmax);
    \draw[->] (argmax) -- (demap) node [midway, above] () {$\hat{a}[k']$};
    \draw[->] (demap) -- (xk) node [pos=0.7,above] {$\hat{\bm{b}}[k']$};

\end{tikzpicture}
        }
        \vspace*{-8mm}
        \caption{Proposed equalizer structure}
        \label{fig:ffeq}
    \end{center}
    \vspace{-1.2cm}
\end{wrapfigure}
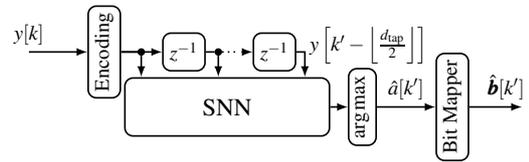
An SNN consists of multiple interconnected state-dependent neurons whose internal states evolve and exhibit temporal dynamics. 
Synapses connect neurons employing adjustable weights.
A common neuron model is the leaky integrate-and-fire (LIF) model \cite{BPTT}:
A neuron's input is scaled by the weight of the connecting synapse.
The neuron's state is charged by integrating the input over time. 
In parallel, it loses charge over time (leakage). 
If the neuron is charged sufficiently, such that its neuronal threshold is exceeded, the neuron fires an output spike, which is passed to the connected upstream neurons. 
In this work, we simulate SNNs using $\texttt{Norse}$ \cite{norse}, which uses the backpropagation through time algorithm \cite{BPTT} to update the synapses' weights during training.
Figure 1 outlines the structure of the SNN-based equalizer as proposed by \cite{arnold_soft}.
The most recent channel output, $y[k]$, is encoded and fed to the equalizer to estimate the $k'$-th transmitted bit pattern $\bm{b}[k']$ corresponding to the $k'$-th transmit symbol, where $k'=k-\left\lceil d_\text{tap}/2\right\rceil$.
The number of equalizer taps $d_\text{tap}$ matches the number of significant channel taps when sampling the channel at the symbol rate.
The encoded sample and previous encoded samples are fed to the SNN. 
At the SNN's output layer, the index $\hat{a}$ of the highest state output neuron is determined and transformed to bits via a bit mapper.

\vspace*{-1mm}
\section{Encoding}
\vspace*{-2mm}

In Fig. 2, we outline the structure of a parameterized encoding, whose parameters are jointly optimized with the loss function of the SNN's learning task.
For encoding, $y\in \mathbb{R}$ is quantized by a uniform quantizer $Q_N(\cdot)$ with $N$ quantization levels and mapped to a class $n$, indicating the quantized value.
Matrices $\bm{W}^{(n)} \in \mathbb{R}^{M \times T}$ are initialized with random i.i.d. elements $w^{(n)}_{m,t}~\sim~\mathcal{N}(0,1)$, where $m\in\{1,\ldots M\}$ denotes the SNN's input neuron, $t~\in~\{1,\ldots T\}$ the SNN's discrete simulation time step and $n\in\{1,\ldots N\}$ the input value class.
Depending on $n$, the matrix $\bm{W}^{(n)}$ is chosen, where the $m$-th row of $\bm{W}^{(n)}$, denoted as $\bm{w}^{(n)}_m$, is fed to the $m$-th SNN's input neuron over time. 
During training, both the matrices $\bm{W}^{(n)}$ and the SNN parameters can be jointly optimized by minimizing the cross entropy loss $J_\mathrm{CE}(a,\hat{a})$ of the transmit symbol's actual index~$a$ and the estimated index $\hat{a}$. 

Sparsity describes how much a signal's energy is concentrated on a few samples \cite{Cherni20}.
To reduce the number of spikes and, therefore, the network's energy consumption, we propose incorporating a sparsity-inducing penalty for $\bm{W}^{(n)}$.
The average $\ell_p$-over-$\ell_q$ quasinorm-ratio over all classes provides a measure of sparseness.
In particular, the normalized $\ell_1$-over-$\ell_2$ quasinorm-ratio $\ell_{1,2}\left(\bm{W}^{(n)}\right)~=~ \sum_{m=1}^{M} \sum_{t=1}^{T} |w^{(n)}_{m,t}| \left( \sum_{m'=1}^{M} \sum_{t'=1}^{T} |w^{(n)}_{m',t'}|^2\right)^{-1/2}$ is frequently used \cite{Cherni20}.
The overall loss function can be defined as ${\left(a,\hat{a},\bm{W}^{(n)}\right)=(1-\alpha)J_\mathrm{CE}(a,\hat{a})+\alpha \frac{1}{N} \sum_{n=1}^N \ell_{1,2}\left(\bm{W}^{(n)}\right), \; \alpha \in [0,1]}$.
To avoid exploding parameters, the matrices $\bm{W}^{(n)}$ are normalized according to $\bm{W}^{(n)}~\leftarrow~\bm{W}^{(n)}\left(\text{max}_{m,t}~\;~w^{(n)}_{m,t}\right)^{-1},~\forall~n\in N$, after each optimization step.
Consequently, low values $w_{m,t}^{(n)}$ are pushed even further to zero.

The proposed encoding is limited by the real-valued nature of the sequences $\bm{w}^{(n)}_m$, losing the binary character of log-scale and ternary encoding.
However, Intel's Loihi 2 chip, as state-of-the-art neuromorphic hardware, supports the input of quantized values, so-called \emph{graded spikes}, with up to 32-bit resolution \cite{Loihi}.
The setup of Fig. 2 can be extended by quantizers $Q(\cdot)$ to simulate the impact of quantization of~$\bm{w}^{(n)}_m$.

\begin{figure}
    \vspace*{-1.2cm}
    \begin{center}
    \resizebox{0.7\textwidth}{!}{
    \begin{tikzpicture}[>=latex, thick]
    \def \xdist{0.5cm}

    \node[] (r) {$y$};
    \node[right=\xdist of r, draw, rectangle, rounded corners] (class) {$Q_N: \mathbb{R} \rightarrow \{1,\ldots N\}$};
    \node[right=\xdist of class,draw, rectangle, rounded corners] (W_emb) {\begin{tikzpicture}[>=latex,thick]
    \def\dist{2cm}

    \node[minimum height=0pt, minimum width=0pt] (base) {};
    \node[above=\dist of base,minimum height=0pt] (y_max) {};
    \draw[->, sharp corners] (base) -| (y_max) node[pos=0.95, right=0cm] {neuron};
    \node[right=\dist of base, minimum width=0pt] (x_max) {};
    \draw[->, sharp corners] (base) |- (x_max) node[pos=0.95, below=0cm] {time};
    \node[above right=\dist of base,minimum width=0pt,minimum height=0pt] (z_max) {};
    \draw[->] (base)+(-0.2pt,-0.2pt) -- (z_max) node[pos=0.99, below right=-1pt and -3pt] {class};

    \node[above right = 0.5pt and 0.5pt of x_max, draw, rectangle, sharp corners, minimum width=3cm, minimum height=2cm] (m1) {$\begin{matrix} w^{(1)}_{1,1} & \ldots & w^{(1)}_{1,T} \\ \vdots & \ddots & \vdots\\ w^{(1)}_{M,1} & \hdots& w^{(1)}_{M,T} \end{matrix}$};

    \node[above right = -1.2cm and -1.8cm of m1,draw, rectangle, sharp corners, minimum width=3cm, minimum height=2cm] (m2) {$\begin{matrix} w^{(2)}_{1,1} & \ldots & w^{(2)}_{1,T} \\ \vdots & \ddots & \vdots\\ w^{(2)}_{M,1} & \hdots& w^{(2)}_{M,T} \end{matrix}$};
    
    \node[above right = 0.0cm and 0.0cm of m1,draw, rectangle, sharp corners, minimum width=3cm, minimum height=2cm] (m3) {$\begin{matrix} w^{(N)}_{1,1} & \ldots & w^{(N)}_{1,T} \\ \vdots & \ddots & \vdots\\ w^{(N)}_{M,1} & \hdots& w^{(N)}_{M,T} \end{matrix}$};
    \draw[dashed] (m1.north west) -- (m3.north west);
    \draw[dashed] (m1.south east) -- (m3.south east);

    \node[above right = -1.2cm and -1.8cm of m1,draw, rectangle, sharp corners, minimum width=3cm, minimum height=2cm, fill=white] (m2) {$\begin{matrix} w^{(2)}_{1,1} & \ldots & w^{(2)}_{1,T} \\ \vdots & \ddots & \vdots\\ w^{(2)}_{M,1} & \hdots& w^{(2)}_{M,T} \end{matrix}$};
    \node[above right = 0.5pt and 0.5pt of x_max, draw, rectangle, sharp corners, minimum width=3cm, minimum height=2cm,fill=white] (m1) {$\begin{matrix} w^{(1)}_{1,1} & \ldots & w^{(1)}_{1,T} \\ \vdots & \ddots & \vdots\\ w^{(1)}_{M,1} & \hdots& w^{(1)}_{M,T} \end{matrix}$};
\end{tikzpicture}};
    \draw[->] (r) -- (class);
    \draw[->] (class) -- (W_emb) node [pos=0.4, above] {$n$};

    \node[above right=-1.3cm and 2*\xdist of W_emb,draw, rectangle, rounded corners, dotted] (Q1) {$Q(\cdot)$};
    \node[above right= -2.25cm and 2*\xdist of W_emb,draw, rectangle, rounded corners, dotted] (Q2) {$Q(\cdot)$};
    \node[below right=-1.5cm and 2*\xdist of W_emb,draw, rectangle, rounded corners, dotted] (Q3) {$Q(\cdot)$};
    \draw[->] (W_emb.east) |- (Q1) node [pos=0.75, above] () {$\bm{w}^{(n)}_1$};
    \draw[->] (W_emb.east) |- (Q2) node [pos=0.75, above] () {$\bm{w}^{(n)}_2$};
    \draw[->] (W_emb.east) |- (Q3) node [pos=0.75, above] () {$\bm{w}^{(n)}_M$};
    \draw[white] (Q2) -- (Q3) node[midway,black] () {$\vdots$};

    \node[right=\xdist of Q1, draw, circle, inner sep=5pt] (in1) {};
    \node[right=\xdist of Q2, draw, circle, inner sep=5pt] (in2) {};
    \node[right=\xdist of Q3, draw, circle, inner sep=5pt] (in3) {};
    \draw[->] (Q1) -- (in1);
    \draw[->] (Q2) -- (in2);
    \draw[->] (Q3) -- (in3);

\end{tikzpicture}
    }
    \caption{Sketch of the embedding. The received value $y$ is classified into a class $n$ with $n\in\{1,\ldots N\}$. Depending on $n$, the embedding matrix $\bm{W}^{(n)}~\in~\mathbb{R}^{M~\times~T}$ is chosen, where $w^{(n)}_{m,t}$ is the embedding value of class $n$, fed to the $m$-th input neuron at time step $t$. To simulate quantized input to neuromorphic hardware, quantizers $Q(\cdot)$ can be included.}
    \label{fig:Embedding}
    \end{center}
    \vspace*{-1.2cm}
\end{figure}
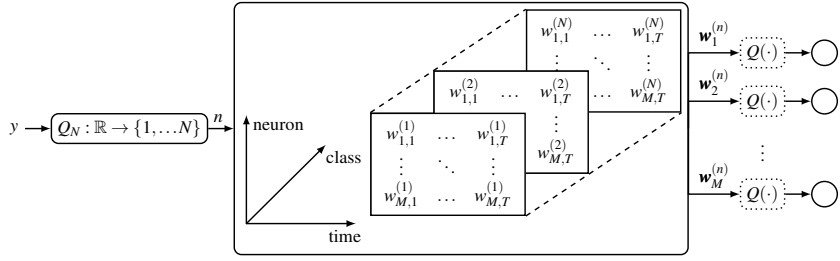
\vspace*{-1mm}
\section{Results}
\vspace*{-2mm}
We compare the proposed encoding with ternary encoding \cite{bansbach} and log-scale encoding \cite{arnold_soft} for an IM/DD link and the SNN-based equalizer of Fig. 1.
The IM/DD link is simulated as in \cite{arnold_soft} with parameters like channel B of \cite{vonBank}: A single mode fiber of $\SI{5}{\kilo\meter}$, RRC pulse shaping with roll-off factor $\beta=0.2$, $100 \, \text{GBd}$ baud rate, wavelength of $\SI{1270}{\nano \meter}$, dispersion coefficient of $-17\,\text{ps}\,\text{nm}^{-1}\,\text{km}^{-1}$ and  $d_\text{tap}=41$ equalizer taps are used in the simulation. 
The pulse amplitude modulated transmit symbols are taken from the set $\mathcal{C}=\left\{0,1,\sqrt{2},\sqrt{3}\right\}$ with Gray mapping. After pulse shaping, a bias is added.
For the log-scale encoding \cite{arnold_soft}, an input neuron count of $M=10$ was used with $T=30$ discrete time steps.
In contrast, ternary encoding \cite{bansbach} and the proposed encoding have an input neuron count of $M=8$ and $T=10$ discrete time steps. 
Furthermore, the proposed encoding uses $N=256$ matrices $\bm{W}^{(n)}$.
Each SNN consists of $N_\text{o}=4$ output neurons and $N_\text{h}=80$ hidden neurons.
The training was carried out using a batch size of $200\,000$, $5$ Epochs, $2000$ batches per epoch, and a learning rate of $10^{-3}$. 
In \cite{vonBank}, we have shown that for the given link, SNNs trained at a noise power of $\sigma^2=-\SI{17}{\decibel}$ perform best. Hence, we fix $\sigma^2=-\SI{17}{\decibel}$ for training.

Figure 3 shows the performance of the proposed encoding and benchmarks. 
The different approaches are compared regarding their equalization performance (BER) and energy efficiency, measured by their respective spike rate.
The spike rate is defined by $S_\mathrm{h}/(N_\text{h}T)$, where $S_\mathrm{h}$ is the number of spikes generated by the hidden layer, and $N_\text{h}T$ is the number of possible spikes of the discrete-time SNN simulation.
Figure 3(a) compares the proposed encoding with the benchmarks for different $\alpha$.
If $\alpha$ is sufficiently low, our encoding is superior to log-scale and ternary encoding and enhances system performance.
\begin{minipage}[t]{1\textwidth}
    \begin{minipage}[]{0.7\textwidth}
        \begin{center}
            \resizebox{!}{5cm}{
                \pgfplotsset{
layers/my layer set/.define layer set={
background,
main,
up
}{
 },
    set layers=my layer set,
}

\begin{tikzpicture}[spy using outlines={rectangle, magnification=2.7, size=1cm, connect spies}]
    \def\lwidth{1.5}
    \def\opac{50}
    \def\marksz{2pt}

    \def\spywidth{2.0cm}
    \def\spyheigth{1.5cm}

    \begin{axis}[
        ymode = log,
        xscale=1,
        xlabel style = {align=center},
        xlabel=$\sigma^2$ (dB) \\ (a),
        x label style={at={(axis description cs:0.5,0.04)},anchor=north},
        ylabel=BER,
        y label style={at={(axis description cs:0.03,0.4)},anchor=west},
        xtick = {16,18,20,22},
        xticklabels={$-16$,$-18$,$-20$,$-22$},
        grid=major,
        legend cell align={left},
        legend style={
            at={(0.02,0.01)},
            anchor=south west,
            fill opacity = 0.8,
            draw opacity = 1, 
            text opacity = 1,
        },
        xmin=15,xmax=23,
        ymin=5e-6,ymax=6e-2,
        axis line style=thick,
        tick label style={/pgf/number format/fixed},
        ]

        \coordinate (spypoint) at (axis cs:21.1,4.5e-4); %
        \coordinate (spyviewer) at (axis cs:21.6,7e-3); %
        \draw [fill=white] ($(spyviewer)+(1cm,0.8cm)$) rectangle ($(spyviewer)-(1cm,0.8cm)$);
        
        \spy[width=2cm,height=1.6cm, every spy on node/.append style={ultra thin},thin,line width=0.01, spy connection path={
        \draw [opacity=0.5] (tikzspyonnode.south west) -- (tikzspyinnode.south west);
        \draw [opacity=0.5] (tikzspyonnode.south east) -- (tikzspyinnode.south east);
        \draw [opacity=0.5] (tikzspyonnode.north west) -- (tikzspyinnode.north west);
        \draw [opacity=0.5] (tikzspyonnode.north east) -- (intersection of  tikzspyinnode.north east--tikzspyonnode.north east and tikzspyinnode.south east--tikzspyinnode.south west);
        ;}] on (spypoint) in node at (spyviewer); %

        \addplot[mark=none, color=black!\opac!white,line width=\lwidth,dashed, opacity= \opac] table [x = sigma, y = Arnold]{./figures/BER_Alpha.txt};
        \addplot[mark=none,every mark/.append style={solid, fill=black!\opac!white}, color=black!\opac!white,line width=\lwidth, opacity =\opac] table [x = sigma, y = Ternary]{./figures/BER_Alpha.txt};
    
        \addplot[dotted, mark=none,mark size = \marksz,every mark/.append style={solid, fill=KITorange}, color=KITpurple, line width=\lwidth] table [x=sigma, y=1e2]{./figures/BER_Alpha.txt};
        \addplot[dashed,mark=none,mark size = \marksz,every mark/.append style={solid, fill=KITorange}, color=KITpurple, line width=\lwidth] table [x=sigma, y=581e4]{./figures/BER_Alpha.txt};
        \addplot[mark=none,mark size = \marksz, color=KITpurple,line width=\lwidth] table [x = sigma, y = 1e9]{./figures/BER_Alpha.txt};

        \addlegendimage{line width=1pt, dashed, black!\opac!white} \addlegendentry{\textit{Log-scale}}
        \addlegendimage{mark=triangle*,every mark/.append style={solid, fill=black!\opac!white},line width=1pt, solid, black!\opac!white} \addlegendentry{\textit{Ternary}}
        
        \addlegendimage{mark=diamond*,every mark/.append style={solid, fill=KITcyanblue!\opac!white},line width=1pt, dashed, KITcyanblue!\opac!white} \addlegendentry{$\alpha =10^{-2}$}
        \addlegendimage{mark=square,every mark/.append style={fill=KITcyanblue!\opac!white,color=KITcyanblue!\opac!white},line width=1pt, dashed, KITcyanblue!\opac!white} \addlegendentry{$\alpha =5.8\cdot 10^{-4}$}
        \addlegendimage{line width=1pt,solid,KITpurple} \addlegendentry{$\alpha =10^{-9}$}

    \begin{pgfonlayer}{up}
        
    \end{pgfonlayer}
\end{axis}
\end{tikzpicture}
            }
            \hspace{-5mm}
            \resizebox{!}{5cm}{
                \pgfplotsset{
layers/my layer set/.define layer set={
background,
main,
up
}{
 },
    set layers=my layer set,
}

\begin{tikzpicture}[spy using outlines={rectangle, magnification=2.7, size=1cm, connect spies}]
    \def\lwidth{1.5}
    \def\opac{50}
    \def\marksz{2pt}

    \def\spywidth{2.0cm}
    \def\spyheigth{1.5cm}

    \begin{axis}[
        ymode = log,
        xscale=1,
        xlabel style = {align=center},
        xlabel=$\sigma^2$ (dB) \\(b),
        x label style={at={(axis description cs:0.5,0.04)},anchor=north},
        yticklabels={,,},
        xtick = {16,18,20,22},
        xticklabels={$-16$,$-18$,$-20$,$-22$},
        grid=major,
        legend cell align={left},
        legend style={
            at={(0.02,0.01)},
            anchor=south west,
            fill opacity = 0.8,
            draw opacity = 1, 
            text opacity = 1,
        },
        xmin=15,xmax=23,
        ymin=5e-6,ymax=6e-2,
        axis line style=thick,
        tick label style={/pgf/number format/fixed},
        ]

        \coordinate (spypoint) at (axis cs:20,5.5e-4); %
        \coordinate (spyviewer) at (axis cs:21.6,7e-3); %
        \draw [fill=white] ($(spyviewer)+(1cm,0.8cm)$) rectangle ($(spyviewer)-(1cm,0.8cm)$);
        
        \spy[width=2cm,height=1.6cm, every spy on node/.append style={ultra thin},thin,line width=0.01, spy connection path={
        \draw [opacity=0.5] (tikzspyonnode.south west) -- (tikzspyinnode.south west);
        \draw [opacity=0.5] (tikzspyonnode.south east) -- (tikzspyinnode.south east);
        \draw [opacity=0.5] (tikzspyonnode.north west) -- (tikzspyinnode.north west);
        \draw [opacity=0.5] (tikzspyonnode.north east) -- (intersection of  tikzspyinnode.north east--tikzspyonnode.north east and tikzspyinnode.south east--tikzspyinnode.south west);
        ;}] on (spypoint) in node at (spyviewer); %

        \addplot[mark=none, color=black!\opac!white,line width=\lwidth,dashed, opacity= \opac] table [x = sigma, y = Arnold]{./figures/BER_Quant.txt};
        \addplot[mark=none,every mark/.append style={solid, fill=black!\opac!white}, color=black!\opac!white,line width=\lwidth, opacity =\opac] table [x = sigma, y = Ternary]{./figures/BER_Quant.txt};

        \addplot[solid,mark=none,mark size = \marksz, color=KITpurple,line width=\lwidth] table [x = sigma, y = float]{./figures/BER_Quant.txt};
        
        \addplot[solid,mark=none,mark size = \marksz,every mark/.append style={solid, fill=KITcyan!white}, color=KITpalegreen, line width=\lwidth] table [x=sigma, y=w4]{./figures/BER_Quant.txt};
        \addplot[dashed, mark=none,mark size = \marksz,every mark/.append style={solid, fill=KITorange!\opac!white}, color=KITpalegreen, line width=\lwidth] table [x=sigma, y=w6]{./figures/BER_Quant.txt};
        \addplot[dash dot,mark=none,mark size = \marksz, color=KITpalegreen,line width=\lwidth] table [x = sigma, y = w8]{./figures/BER_Quant.txt};

        \addlegendimage{line width=1pt, dashed, black!\opac!white} \addlegendentry{\textit{Log-scale}}
        \addlegendimage{mark=triangle*,every mark/.append style={solid, fill=black!\opac!white},line width=1pt, solid, black!\opac!white} \addlegendentry{\textit{Ternary}}
        
        \addlegendimage{solid,mark=diamond*,mark size = \marksz, color=KITpurple,line width=\lwidth} \addlegendentry{Float}
        
        \addlegendimage{mark=diamond*,every mark/.append style={solid, fill=KITcyanblue!\opac!white},line width=1pt, dashed, KITcyanblue!\opac!white} \addlegendentry{4 Bit Quantized}
        \addlegendimage{mark=square,every mark/.append style={fill=KITcyanblue!\opac!white,color=KITcyanblue!\opac!white},line width=1pt, dashed, KITcyanblue!\opac!white} \addlegendentry{6 Bit Quantized}
        \addlegendimage{line width=1pt,solid,KITpurple} \addlegendentry{8 Bit Quantized}

    \begin{pgfonlayer}{up}
        
    \end{pgfonlayer}
\end{axis}
\end{tikzpicture}
            }%
        \end{center}
    \end{minipage}%
    \hfill
    \begin{minipage}[]{0.29\textwidth}
    \setlength\tabcolsep{1.5pt}
    \resizebox{!}{2cm}{
        \begin{tabular}{ l c } 
            \toprule
            Encoding & spike rate ($\%$) \\
            \midrule
            \textit{Log-scale} & $14.5$  \\  
            \textit{Ternary} & $12.5$ \\ [0.25em]
            $\alpha=10^{-2}$ & $5.2$ \\
            $\alpha=5.8\cdot10^{-4}$ & $6.0$ \\
            $\alpha=10^{-9}$ & $7.3$ \\[0.25em] 
            4 Bit Quant. & $7.6$ \\
            6 Bit Quant. & $7.3$ \\
            8 Bit Quant. & $7.3$ \\
            \bottomrule
        \end{tabular}}
        \vspace*{4.5mm}
        \begin{center}
            \footnotesize(c)
        \end{center}
        
    \end{minipage}
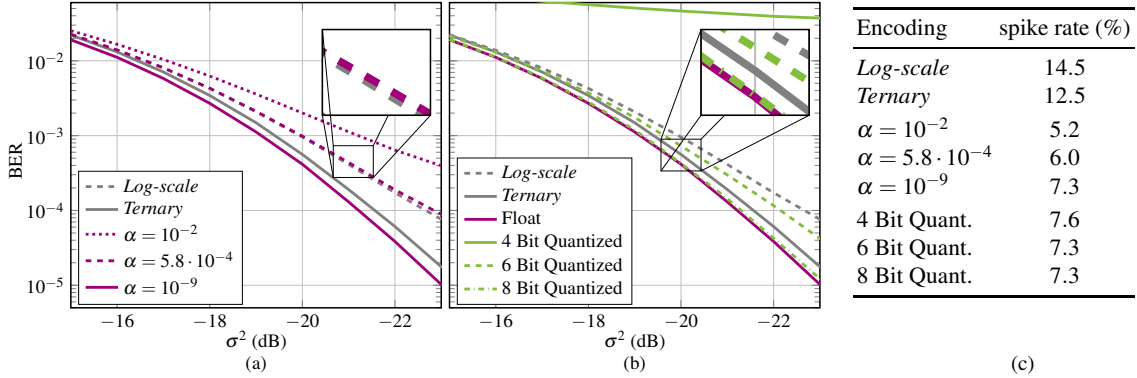
        \footnotesize Fig. 3: Comparison of the proposed encoding against the benchmark encodings: (a) The impact of $\alpha$, (b) the impact of the quantization of $\bm{W}{(n)}$ for fixed $\alpha=10^{-9}$ and (c) the resulting spike rate measured at fixed $\sigma^2=\SI{19}{\decibel}$. 
    \label{fig:results}
\end{minipage}
\vspace*{1mm}
\setcounter{figure}{3}
\begin{wrapfigure}{r}{0.38\textwidth}
    \vspace{-0.9cm}
    \begin{center}
        \resizebox{0.38\textwidth}{!}{
            \pgfplotstableread{
Histogramm.tex
}\pdpfdata

\pgfplotsset{
layers/my layer set/.define layer set={
background,
main,
up
}{
 },
    set layers=my layer set,
}

\begin{tikzpicture}[spy using outlines={rectangle, magnification=2.7, size=1cm, connect spies}]
    \def\lwidth{2.0}
    \def\opac{100}
    \def\marksz{2pt}

    \def\spywidth{2.0cm}
    \def\spyheigth{1.5cm}

    \begin{axis}[
        ymode = log,
        xscale=1,
        xlabel=$w^{(n)}_{m,t}$,
        x label style={at={(axis description cs:0.5,0.04)},anchor=north},
        ylabel=rel. freq.,
        y label style={at={(axis description cs:0.03,0.4)},anchor=west},
        xtick = {-4,-3,-2,-1,0,1,2,3,4},
        xticklabels={$-4$,$-3$,$-2$,$-1$,$0$,$1$,$2$,$3$,$4$},
        grid=major,
        legend cell align={left},
        legend style={
            at={(0.02,0.98)},
            anchor=north west,
            fill opacity = 0.8,
            draw opacity = 1, 
            text opacity = 1,
        },
        xmin=-4,xmax=4,
        ymin=5e-5,ymax=1,
        axis line style=thick,
        tick label style={/pgf/number format/fixed},
        ]

        \addplot[mark=none, color=black!70!white,line width=\lwidth, opacity= 0.7] table [x = x, y = Pre]{./figures/Histogramm.txt};
        \addplot[mark=none, color=KITblue!\opac!white,line width=\lwidth, opacity =1.0] table [x = x, y = Pos1e2]{./figures/Histogramm.txt};
        \addplot[mark=none, color=KITred!\opac!white,line width=\lwidth, opacity=0.4] table [x = x, y = Pos1e9]{./figures/Histogramm.txt};
    
        \addlegendimage{line width=1pt, dashed, blue!\opac!white} \addlegendentry{No opt.}
        \addlegendimage{mark=triangle*,every mark/.append style={solid, fill=red!\opac!white},line width=1pt, solid, red!\opac!white} \addlegendentry{$\alpha=10^{-2}$}
        \addlegendimage{mark=triangle*,every mark/.append style={solid, fill=yellow!50!white,draw opacity=0.5},line width=1pt, solid, yellow!50!white,draw opacity=0.5} \addlegendentry{$\alpha=10^{-9}$}
    
    \begin{pgfonlayer}{up}
        
    \end{pgfonlayer}
\end{axis}
\end{tikzpicture}
        }
        \vspace*{-8mm}
        \caption{Histogram of the elements of $\bm{W}$ before optimization (No opt) and with optimization using $\alpha$.}
        \label{fig:Histogramm}
    \end{center}
    \vspace{-1.2cm}
\end{wrapfigure}
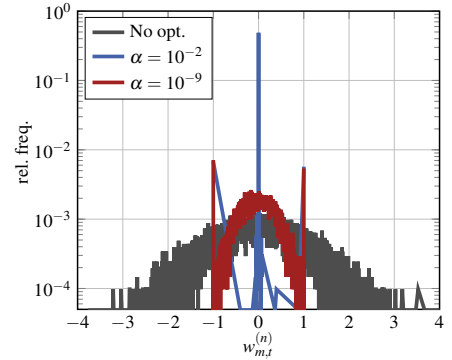
Furthermore, the spike rate is significantly reduced.
For $\alpha=10^{-9}$, it is reduced by roughly $50\%$ compared to the log-scale and roughly $42\%$ compared to the ternary encoding, see Fig. 3(c).
Over the range of simulated $\sigma^2$, the spike rate is near constant.
The spike rate can be further decreased by increasing $\alpha$ and thus the impact of the $\ell_1$-over-$\ell_2$ penalty on the loss function.
However, this rate reduction comes at the cost of decreasing system performance.
For a given target BER, a suitable choice of $\alpha$ enables the flexible reduction of the spike rate. 
Figure 4 shows the impact of $\alpha$ on the distribution of $w_{m,t}^{(n)}$.
Prior to learning, the parameters are initialized by sampling independently from $\mathcal{N}(0,1)$.
For both $\alpha=10^{-2}$ and $\alpha=10^{-9}$, approximately $1.6\%$ of the SNN's input has the maximal amplitude of one.
The number of graded spikes reduces by increasing $\alpha$, which in turn increases the BER.
The histograms and the performance of $\alpha=10^{-2}$ and $\alpha=10^{-9}$ indicate that most input information is encoded in the graded spikes of $\bm{W}^{(n)}$.
Hence, quantizing the encoding values decreases performance, as shown in Fig. 3(b).
If the number of quantization bits is sufficiently high, the quantization has only a minor effect on the system's performance.
However, if the number of quantization bits is set too low, the system performance will deteriorate considerably.
As mentioned above, Intel's Loihi 2 supports up to 32-bit quantization. Hence, the quantization effect can be neglected.
Notably, the spike rate is only negligibly affected by quantization.

\vspace*{-1mm}
\section{Conclusion}
\vspace*{-2mm}
We proposed a novel neural encoding based on learnable parameter matrices.
We have shown that the encoding enhances system performance while decreasing the spike rate significantly.

\end{document}